\documentclass[11pt]{pjm} 
\usepackage{amsthm}
\usepackage{amsfonts}
\usepackage{amsmath}
\usepackage{eucal}

\renewcommand{\labelenumi}{\textup{\theenumi.}}

\newcommand{\tenz}{\ensuremath{T_{ij}^{kl}}}

\newcommand{\hb}{\ensuremath{\mathcal{H}}}

\newcommand{\img}{\operatorname{ran}}
\newcommand{\capker}{\ker}
\newcommand{\id}{\operatorname{id}}
\newlength{\qedskip}

\title[Braided operator commutators]
{The kernel of Fock representations of Wick
algebras with braided operator of coefficients} 

\author[P. E. T. J{\o}rgensen]{Palle E. T. J{\o}rgensen} 
\address{
\textup{(P. J.)}
Department of Mathematics\\The University of Iowa\\
Iowa City, Iowa 52242-1419 U.S.A.} 
\email{jorgen@math.uiowa.edu} 
\thanks{P. J. was partially supported by the NSF}

\author[D. P. Proskurin]{Daniil P. Proskurin} 
\address{
\textup{(D. P. and Yu. S.)}
Institute of Mathematics\\Ukrain\-i\-an National Academy of
Sciences\\
Tereshchinkivs'ka, 3, Kiev, 252601, Ukraine} 
\email{prosk@imath.kiev.ua} 

\author[Yu.
S. Samo\u\i{}lenko]{Yuri\u\i{}~S.~Samo\u\i{}lenko} 
\email{Yurii\_Sam@imath.kiev.ua} 
\thanks{Yu. S. was partially supported by CRDF grant
no. UM1-311}

\subjclass{46L55, 47C05, 81S05, 81T05}
\keywords{unbounded operators,
Hilbert space,
quantum deformation,
positive definite forms}

\theoremstyle{plain}
\newtheorem{definition}{Definition}
\newtheorem{proposition}{Proposition}
\newtheorem{theorem}{Theorem}

\theoremstyle{remark}
\newtheorem{remark}{\bf Remark}
\newtheorem{example}{\bf Example}

\date{Received date / Revised version date} 

\begin {document}

\begin{abstract}
It is shown that the kernel of
the
Fock representation of
a certain
Wick algebra
with braided operator of coefficients $T$, $||T\,||\le 1$, coincides
with the largest quadratic Wick ideal. Improved conditions on the
operator $T$ for the Fock inner product to be strictly positive
are given.
\end{abstract}

\maketitle

\section{Introduction}\label{intro} 
  
The problem of positivity of
the
Fock
space
inner product is
central in the
study of the Fock representation of Wick algebras (see 
\cite{bib:2}, \cite{bib:7},
\cite{bib:8}, \cite{bib:1}).
The paper \cite{bib:1}
presents
several conditions on the coefficients of
the
Wick algebra
for the Fock inner product to be positive. If the operator of
coefficients of
the
Wick algebra $T$ satisfies the braid condition and the
norm restriction $\Vert T\Vert \le 1$, then, as proved in \cite{bib:2},
the Fock inner product is positive.
Moreover if $-1< T <1$, it
was shown
in \cite{bib:2}
that
the
Fock inner product is strictly positive. In this
article we prove that,
for braided $T$ with $\Vert T\Vert\le 1$,
the 
kernel of
the
Fock inner product coincides with the largest quadratic Wick ideal.
In particular this implies that,
for $-1< T\le 1$,
the Fock inner product is
strictly positive
definite,
and the Fock representation is faithful. 

This article is organized as follows. In sec.~\ref{S1} we present
definitions
of Wick algebras and
the
Fock representation and show that, in the braided
case, the kernel of
the
Fock representation is generated by the kernel of
the
Fock inner product. In sec.~\ref{S2} we prove that if
the
operator $T$ is braided
and $\Vert T\Vert\le 1$, then the kernel of
the
Fock inner product coincides
with the two-sided ideal generated by $\ker (1+T)$. In sec.~\ref{S3}
we combine
results obtained in sec.~\ref{S1} and sec.~\ref{S2} to examine
the $C^*$-representability of
certain Wick algebras or their quotients. All results are illustrated
by examples of different kinds of $q_{ij}$-CCR.
\section{Preliminaries}\label{S1}
For more detailed information about Wick algebras and
the
Fock
representation we
refer
the reader to \cite{bib:1}. In this section we present
only the basic definitions and properties.
\medskip

\noindent
{\bf 1.} The notion of a
$*$-algebra allowing Wick ordering (Wick algebra) was
presented in the paper \cite{bib:1} as a generalization of a wide class of
$*$-algebras,
including
the twisted CCR and CAR algebras (see \cite{bib:3}),
the
$q$-CCR (see \cite{bib:5})
algebra, etc.
\begin {definition}
Let  $ \mathbb{J}=\mathbb{J}_{d}
        = \{1,2,\dots ,d\}$, $ \tenz\in C$,
$i, j, k, l\in\mathbb{J}$,
be
such that
$\tenz=\overline{T_{ji}^{lk}}$.
The
Wick algebra with the set of coefficients
$\{T_{ij}^{kl}\}$
is
denoted
$W(T)$,
and
is a $*$-algebra, defined by
generators $a_{i}$, $a_{i}^{*}$, $i\in\mathbb{J}$,
which satisfy the
basic relations:
\[
a_i^*a_j=\delta_{ij}1+\sum_{k,l=1}^{d} \tenz a_{l}a_{k}^{*}
\]
\end {definition}
\begin{definition}
Monomials of the form $a_{i_1}a_{i_2}\cdots a_{i_m}a_{j_1}^*a_{j_2}^*
\cdots a_{j_k}^*$ are called Wick ordered monomials.
\end{definition}
It was proved in \cite{bib:1} that
the
Wick ordered monomials form
a basis for
$W(T)$.

Let $\hb=
\left\langle
e_{1}, \dots ,e_{d} 
\right\rangle
 $. Consider the full tensor algebra over
$\hb,\, \hb^*$, denoted by
$\mathcal{T}(\hb,\hb^{*})$. Then
    \[
W(T) \simeq \mathcal{T}(\hb,\hb^{*}) \biggm/
\left\langle
e_{i}^{*}\otimes e_{j} -
\delta_{ij}1 -\sum \tenz e_{l}\otimes e_{k}^{*} 
\right\rangle .
    \]
To study the structure of Wick algebras,
and the structure of the Fock
representation,
it is useful to introduce
the following operators
on $\mathcal{H}^{\otimes\,n}:=
\smash{\underbrace{\mathcal{H}\otimes \cdots\otimes \mathcal{H}}_{n}}$
(see \cite{bib:1}):
\begin{align*}
T\colon & \mathcal{H} \otimes \mathcal{H}\mapsto \mathcal{H}
 \otimes\mathcal{H}, \quad  T e_{k}\otimes e_{l} =
\sum_{i,j} T_{ik}^{lj}e_{i}\otimes e_{j},\ T=T^*,
\\
T_{i}\colon & \mathcal{H}^{\otimes\,n}\mapsto\mathcal{H}^{\otimes\,n},
\quad
T_{i}=\underbrace{1\otimes \cdots\otimes 1}_{i-1}\otimes T
\otimes\underbrace{1\otimes\cdots\otimes 1}_{n-i-1}\,,
\\
R_{n}\colon & \mathcal{H}^{\otimes\,n}\mapsto \mathcal{H}^{\otimes\,n},
\quad
R_{n}=1+T_1+T_1 T_2+\cdots + T_1 T_2\cdots T_{n-1},
\\
P_n\colon & \mathcal{H}^{\otimes\,n}\mapsto \mathcal{H}^{\otimes\,n},
\quad
P_2=R_2,\ P_{n+1}=(1\otimes P_n)R_{n+1}.
\end{align*}

In this article we suppose that the operator $T$ is contractive, i.e.,
$\Vert T\Vert\le 1$, and satisfies the {\it
braid condition}, i.e., on  $\hb^{\otimes\,3}$ the equality
$T_1 T_2 T_1 = T_2 T_1 T_2$ holds. It follows from the definition of
$T_i$ that
then
$T_iT_j=T_jT_i$ if $\left| i-j\right|\ge 2$, and for the braided $T$
one has $T_iT_{i+1}T_i=T_{i+1}T_iT_{i+1}$.
\begin{remark}
These conditions hold for such well-known algebras as
$q_{ij}$-CCR, $\mu$- CCR, $\mu$-CAR (see \cite{bib:1}).
\end{remark}
The Fock representation of a Wick
$*$-algebra is determined by
a
vector
$\Omega$ such that $a_i^*\Omega=0$ for all $i=1,\dots,d$ (see \cite{bib:1}).
\begin{definition}
[The Fock representation]
The representation $\lambda_0$,
acting on the space $\mathcal{T}(\hb)$
by formulas
\begin{align*}
\lambda_0(a_i)e_{i_1}\otimes\cdots\otimes
e_{i_n}& =e_i\otimes e_{i_1}\otimes\cdots\otimes e_{i_n},\quad
n\in\mathbb{N}\cup\{0\},\\
\lambda_0(a_i^*)1&= 0,
\end{align*}
where the action of $\lambda_0(a_i^*)$ on the monomials of degree $n\ge
1$ is determined inductively using the basic relations, is called the
Fock representation.
\end{definition}
Note that
the
Fock representation is not a $*$-representation with
respect to the standard inner product on $\mathcal{T}(\hb)$. However,
it
was proved in \cite{bib:1} that there exists
a
unique Hermitian sesquilinear
form $\bigl<\,\cdot \,,\,\cdot \,\bigr>_0$ on $\mathcal{T}(\hb)$ such that $\lambda_0$
is a $*$-representation on
$(\mathcal{T}(\hb),\bigl<\,\cdot \,,\,\cdot \,\bigr>_0)$. This form is called
the Fock inner product on $\mathcal{T}(\hb)$.

The subspaces $\hb^{\otimes\,n}$,
$\hb^{\otimes\,m}$, $n\ne m$, are orthogonal with respect to
$\bigl<\,\cdot \,,\,\cdot \,\bigr>_0$,
and on $\hb^{\otimes\,n}$ we have the following
formula (see \cite{bib:1}):
\[
\bigl<X,Y\bigr>_0 =\bigl<X, P_n Y\bigr>,\ n\ge 2.
\]
So, the positivity of the Fock inner product is equivalent to the
positivity of operators $P_n$, $n\ge 2$, and
$\mathcal{I}=\bigoplus_{n\ge 2\vphantom{\ge_{\mathstrut}}}\ker P_n$
determines the kernel of the Fock inner product.
It was noted in \cite{bib:1} that
the
Fock representation is the GNS
representation associated with the linear functional
$f$
on a Wick algebra
such that $f(1)=1$ and, for any Wick ordered monomial,
$f(a_{i_1}\cdots a_{i_n}a_{j_1}^*\cdots a_{j_m}^*)=0$. Then for any
$X,Y\in\mathcal{T}(\hb)$ we have (see \cite{bib:1}):
\[
\bigl< X,Y \bigr>_0=f(X^*Y).
\]
\noindent
{\bf 2.} In the following proposition we describe the kernel of
the
Fock
representation of
a
Wick algebra with braided operator $T$ in terms of
the Fock inner product.
\begin{proposition}
Let $W(T)$ be
the
Wick algebra with braided operator $T$, and let the Fock
representation $\lambda_0$ be positive
\textup{(}i.e., the Fock inner product is
positive
definite\textup{).} Then
$\ker \lambda_0 = \mathcal{I}\otimes\mathcal{T}(\hb^*) +
\mathcal{T}(\hb)\otimes\mathcal{I}^*$.
\end{proposition}
\begin{proof}
First, we show that $X\in\ker P_m$ implies $X\in\ker\lambda_0$.
Indeed, let $Y\in\hb^{\otimes\,n}$; then
\[
\lambda_0(X)Y=X\otimes Y.
\]
Note that for braided $T$ we have the following decomposition (see \cite{bib:2}
and sec.~\ref{S2} for more details):
\[
P_{n+m}=P(D_m)(P_m\otimes\mathbf{1}_n),
\]
where
\begin{align*}
& P(D_m)=\widetilde{R}_{n+m}\widetilde{R}_{n+m-1}\cdots
\widetilde{R}_{m+1},\\
&\widetilde{R}_k =1+T_{k-1}+T_{k-2}T_{k-1}+\cdots+
T_1T_2\cdots T_{k-1},\quad k\ge 2.
\end{align*}
Then
\[
P_{n+m}(\lambda_0(X)Y)=P_{n+m}(X\otimes Y)=P(D_m)(P_m X\otimes Y)=0,
\]
and $\lambda_0(X)=0$ on
$\left(\mathcal{T}(\hb)\, ,\, \bigl<\,\cdot \,,\,\cdot \,\bigr>_0\right)$.
Therefore
$\mathcal{I}\subset\ker\lambda_0$, and since $\ker\lambda_0$ is a
$*$-ideal,
\begin{equation}\label{incl}
\mathcal{I}\otimes\mathcal{T}(\hb^*)+\mathcal{T}(\hb)
\otimes\mathcal{I}^*\subset\ker\lambda_0.
\end{equation}

To prove the converse inclusion,
we need a formula determining the
action of $\lambda_0(X^*)$ on
$\mathcal{T}(\hb)$ for any $X\in\hb^{\otimes\,k}$, $k\in\mathbb{N}$.
For $k=1$, $X=e_i$, $i=1,\dots,d$, it was
proved
in \cite{bib:1} that:
\[
\lambda_0(e_i^*) Y=\mu(e_i^*)R_n Y,\quad \forall\, Y\in\hb^{\otimes\,n},
\]
where $\mu(e_i^*)\colon\mathcal{T}(\hb)\mapsto\mathcal{T}(\hb)$ is the
annihilation operator:
\[
\mu(e_i^*)e_{i_1}\otimes e_{i_2}\otimes\cdots\otimes e_{i_n}=
\delta_{ii_1} e_{i_2}\otimes\cdots\otimes e_{i_n} .
\]
Then, using the definition of $P_n$, it is easy to see that, for
$X\in\hb^{\otimes\,n}$ and $Y\in\hb^{\otimes\,n}$,
\[
\lambda_0(X^*)Y=\bigl< X,P_n Y\bigr>=\bigl< X, Y\bigr>_0 .
\]
Let now
\[
Z=\sum_{i=1}^n Y_i X_i^*+\sum_{j=n+1}^l Y_jX_j^*\in\ker\lambda_0,
\]
where $Y_i\in\mathcal{T}(\hb)$, $i=1,\dots,l$,
\[
X_i\in\hb^{\otimes\,{m}},\; i=1,\dots,n\, ,\quad
X_j\in\hb^{\otimes\,{n_j}},\; n_j>m\,,\quad j=n+1,\dots,l.
\]
Now
(\ref{incl}) implies that we
can suppose that
the
elements $X_i$ are linearly independent modulo
$\mathcal{I}$. Denote by 
$\{\widehat{X}_i,\ i=1,\dots,n\}\subset\hb^{\otimes\,m}$
a
family dual to the $\{X_i,\ i=1,\dots,n\}$ with respect to
$\bigl<\, \cdot\,  ,\, \cdot\, \bigr>_0$, i.e., such that
\[
\bigl<X_i\, ,\, P_m\widehat{X}_j\bigr>=
\bigl< X_i\, ,\, \widehat{X}_j\bigr>_0=\delta_{ij},
\quad i,j=1,\dots,n.
\]
Since, for any $j=n+1,\dots,l$ and $i=1,\dots,n$,
\[
\lambda_0(X_j^*)\widehat{X}_i=0,
\]
we have, in $\left(\mathcal{T}(\hb)\, ,\, \bigl<\,\cdot \,,\,\cdot \,\bigr>_0\right)$,
\[
0=\lambda_0(Z)\widehat{X}_i = Y_i,\quad i=1,\dots,n,
\]
which implies $Y_i\in\mathcal{I}$, $i=1,\dots,n$. The proof can be
completed by evident induction.
\end{proof}
\begin{remark}
In particular, we have shown, for braided $T$, and for any
$X\in\hb^{\otimes\,n}$ and $Y\in\ker P_m$, that
\[
X\otimes Y\in\ker P_{n+m}.
\]
By
similar arguments, $Y\otimes X\in\ker P_{n+m}$, i.e.,
$\mathcal{I}=\smash{
\left\langle\bigotimes_{n\ge 2}\ker P_n \right\rangle}$
is a two-sided
ideal in $\mathcal{T}(\mathcal{H})$.
\end{remark}

The two-sided ideal $\mathcal{J}\subset\mathcal{T}(\hb)$ is called
a
Wick ideal
(see \cite{bib:1}) if it satisfies
the following
condition:
\begin{equation}\label{eq:wick}
\mathcal{T}\left(\hb^*\right)\otimes\mathcal{J}
\subset\mathcal{J}\otimes\mathcal{T}\left(\hb^*\right).
\end{equation}
If $\mathcal{J}$ is generated by some subspace of $\hb^{\otimes\,n}$,
then $\mathcal{J}$ is
called
a
homogeneous Wick ideal of degree $n$.

We
show that for Wick algebras with braided operator of
coefficients, $\mathcal{I}$ is a Wick ideal .
\begin{proposition}
Let $T$ satisfy the braid condition, and
$\mathcal{I}=\smash{
\left\langle\bigoplus_{n\ge 2}\ker P_n\right\rangle}$;
then
\begin{equation}\label{eq:sufwick}
\hb^*\otimes\mathcal{I}\subset\mathcal{I}+\mathcal{I}\otimes\hb^*.
\end{equation}
\end{proposition}
\begin{proof}
Note that conditions (\ref{eq:wick}) and (\ref{eq:sufwick}) are
equivalent (see \cite{bib:1}).
To prove
the
proposition, it is sufficient to show that, if $X\in\ker P_n$
for some $n\ge 2$, then for any $i=1,\dots,d$,
\[
e_i^*\otimes X\in\ker P_{n-1}+\ker P_n\otimes\hb^*.
\]
Indeed, for any $X\in\hb^{\otimes\,n}$,
we have the following formula
(see \cite{bib:6}):
\[
e_i^*\otimes X=\mu(e_i^*)R_n X+\mu(e_i^*)\sum_{k=1}^d
T_1T_2\cdots T_n (X\otimes e_k)\otimes e_k^*.
\]
Then for $X\in\ker P_n$,
we have
\[
P_{n-1}\mu(e_i^*)R_n X=\mu(e_i^*)(1\otimes P_{n-1})R_nX=
\mu(e_i^*)P_n X=0.
\]
Note that,
for braided $T$, for any $k=2,\dots,n$,
\[
T_k (T_1T_2\cdots T_n)=(T_1T_2\cdots T_n)T_{k-1},
\]
which implies that
\[
(1\otimes P_n)(T_1T_2\cdots T_n)=(T_1T_2\cdots T_n) (P_n\otimes 1).
\]
Then for any $k=1,\dots,d$,
\begin{align*}
P_n\mu(e_i^*)T_1T_2\cdots T_n (X\otimes e_k)
& =\mu(e_i^*)(1\otimes P_n)T_1T_2\cdots T_n (X\otimes e_k)\\
& =\mu(e_i^*)T_1T_2\cdots T_n (P_n\otimes 1)(X\otimes e_k)=0.
\settowidth{\qedskip}{$\displaystyle
P_n\mu(e_i^*)T_1T_2\cdots T_n (X\otimes e_k)
=\mu(e_i^*)T_1T_2\cdots T_n (P_n\otimes 1)(X\otimes e_k)=0.$}
\addtolength{\qedskip}{-\textwidth}
\rlap{\hskip-0.5\qedskip\llap{\qedsymbol}}
\end{align*}
\renewcommand{\qed}{\relax}
\end{proof}

For Wick algebras with braided $T$,
the largest homogeneous ideal of
degree $n$ is generated by $\ker R_n$ (see \cite{bib:1} and \cite{bib:6}), i.e.,
the
condition $\ker R_n\ne\{0\}$ is necessary and sufficient for the
existence of homogeneous Wick ideals. In the following proposition we
show that the same is true for
arbitrary Wick ideals.
\begin{theorem}\label{prop:idew}
If $\mathcal{J}\subset\mathcal{T}(\hb)$ is
a
non-trivial Wick ideal, then
there exists $n\ge 2$ such that $\ker R_n\ne\{0\}$.
\end{theorem}
\begin{proof}
For any $X\in\mathcal{T}(\hb)$,
by $\deg X$ we denote the highest degree
of its homogeneous components. Let $Y\in\mathcal{J}$ be of minimal
degree.
\[
Y=Y_1+Y_2+\cdots +Y_k,\quad Y_i\in\hb^{\otimes\,{n_i}},\ i=1,\dots,k
,\ n_i\in\mathbb{N}\cup\{0\}.
\]
Suppose that $\deg Y\ge 2$: then for any $i=1,\dots,d$, we have
\[
e_i^*\otimes Y=\sum_{j=1}^k \mu(e_i^*)R_{n_j}Y_j +
\mu(e_i^*)\sum_{j=1}^k\sum_{l=1}^d \widetilde{T}_{n_j}(Y_j\otimes e_l)
\otimes e_l^*,
\]
where we put $R_0=1$, $R_1=1$, and
\[
\widetilde{T}_k=
\begin{cases}
T_1T_2\cdots T_{k},&\quad k\ge 2,\\
T ,&\quad k=1,\\
1,&\quad k=0.
\end{cases}
\]
Then condition (\ref{eq:sufwick}) implies that for any $i=1,\dots,d$,
\[
\sum_{j=1}^k \mu(e_i^*)R_{n_j}Y_j \in\mathcal{J}.
\]
Since the degrees of these elements
are
less than the degree of $Y$,
we conclude that
\[
\sum_{j=1}^k \mu(e_i^*)R_{n_j}Y_j =0,\quad i=1,\dots,d,
\]
and the independence of
the
Wick ordered monomials
then
implies
\[
\mu(e_i^*)R_{n_j}Y_j =0,\quad i=1,\dots,d.
\]
Let $Y_k$ be the highest homogeneous component of $Y$;
then, by our
assumption, $\deg Y_k\ge 2$, and $\sum_{i=1}^d
e_i\mu(e_i^*)R_{n_k}Y_k=R_{n_k}Y_k=0$, i.e., $Y_k\in\ker R_{n_k}$.

To complete the proof, note that if $X=\beta+\sum_{i=1}^d \alpha_i
e_i\in\mathcal{J}$, then for any $j$, we have
\[
e_j^*\otimes X=\alpha_j +\beta e_j^*+\sum_{i=1}^d\alpha_i\sum_{k,l=1}^d
T_{ji}^{kl}e_l\otimes e_k^*,
\]
and (\ref{eq:sufwick}) implies $\alpha_j=0$, $j=1,\dots,d$, $\beta =0$.
\end{proof}

\section{The structure of $\ker P_n$}\label{S2}
In this section, we show that for Wick algebras with braided $T$
satisfying
the
condition
$-1<T\le 1$, the Fock representation is faithful,
and for $-1\le T\le 1$,
the kernel of
the
Fock representation is generated by the largest
quadratic Wick ideal (the largest quadratic Wick ideal is the largest
homogeneous Wick ideal of degree $2$).

To do
this
we need some properties of quasimultiplicative maps on the
Coxeter group $S_n$ (for more detailed information we
refer
the
reader to \cite{bib:2}).
\medskip

\noindent
{\bf 1.} Consider
$S_{n+1}$ as a Coxeter group, i.e., a group defined
as follows:
$
S_{n+1} =
\left\langle
\sigma_i: \sigma_i^2 = e,\ \sigma_i\sigma_j =
\sigma_j\sigma_i,\ \left| i-j\right| \geq 2,\ \sigma_i\sigma_{i+1}\sigma_i
=\sigma_{i+1}\sigma_i\sigma_{i+1},\ i=1,\dots,n
\right\rangle
$.
In order to study the invertibility of $P_n$ for any family of
operators $\{T_i\, ,\ i=1\dots ,n,\
T\in B(\mathcal{K})\}$, satisfying
the
conditions
\[
T_i T_{i+1}T_i = T_{i+1}T_i T_{i+1},\ T_i^* = T_i,\ -1\le T_i\le1,
\]
where $\mathcal{K}$
is a
separable Hilbert space,
we may define (as in \cite{bib:2})
the
function
\[
     \phi\colon S_{n+1}\mapsto B(\mathcal{K})
\]
by
the
formulas
     \begin{gather}
     \phi(e)=1,\ \ \ \phi(\sigma_i)=T_i,\\
     \phi(\pi)=T_{i_1}\cdots T_{i_k},
     \end{gather}
where $\pi = \sigma_{i_1}\cdots\sigma_{i_k}$ is a reduced
decomposition. It was shown in \cite{bib:2} that
\[
P_{n+1} = P(S_{n+1})=\sum_{\sigma\in S_{n+1}}\phi(\sigma).
\]
Denote by $S$ the set of generators of $S_{n+1}$ as a Coxeter
group. Consider, for any $ J\subset S$, the set
     \[
     D_{J} = \{\sigma\in S_{n+1} \mid \left|\sigma s\right| =
\left|\sigma\right| +1,\; \forall\, s\in J\}.
     \]
Let  $W_{J}$ be a Coxeter group, generated by $J$. Then
$S_{n+1} = D_J W_J$ (see \cite{bib:4}), and 
$P_{n+1}=P(D_J)P(W_J)$ (see \cite{bib:2}). Using the equalities
$P_{n+1}^*=P_{n+1}$, 
$P(W_J)^*=P(W_J)$, we obtain $P_{n+1}=P(W_J)P(D_J)^*$,
where for all $M\subset S_{n+1}$,
\[
P(M)=\sum_{\sigma\in M}\phi(\sigma).
\]
In what follows we use a quasimultiplicative analogue of the
Euler-Solomon formula (see \cite[Lemma 2.6]{bib:2}):
\begin{equation}  \label{eq:eul}
     \sum_{\substack{J\subset S\\
         J\neq S,\,J\neq\emptyset}}(-1)^{\left| J\right|}P(D_J)=
     -(-1)^{\left| S\right|}1+\phi(\sigma_0^{(n+1)})-P(S_{n+1}),
\end{equation}
where $\sigma_0^{(n+1)}$ is the unique element of $S_{n+1}$ with maximal 
possible length of the reduced decomposition.
\begin{remark}
\begin{enumerate}
\item
The element $\sigma_0^{(n+1)}$
of
the group $S_{n+1}$ has the form
     \[
     \sigma_0^{(n+1)}=
(\sigma_1\cdots\sigma_n)(\sigma_1\cdots\sigma_{n-1})\cdots
(\sigma_1\sigma_2)\sigma_1.
     \]
Set
$U_n=\phi(\sigma_0^{(n+1)})$: then
\[
U_n=(T_1T_2\cdots T_n)(T_1T_2\cdots T_{n-1})\cdots
(T_1T_2)T_1.
     \]
\item
It is easy to see that
the
operator $U_n$ is selfadjoint, and,
taking adjoints,
we can rewrite (\ref{eq:eul}) in the
following form:
     \begin{equation}\label{adeul}
     \sum_{J\subset S,\,J\neq\emptyset}(-1)^{\left| J\right|}P(D_J)^*=
     (-1)^{n+1}1+U_n-P_{n+1}.
     \end{equation}
\item
Note also that, for all $J\subset S$,
the group $W_J$ is isomorphic
to
$S_k$ for some $ k<n$, or to the direct product of
some
such  groups.
\end{enumerate}
\end{remark}
\noindent
{\bf 2.} In what follows we shall use the following properties of the 
operator $U_n$.
\begin{proposition}
$\ker P_{n+1}$ is invariant with respect to the action of
$U_n$.
\end{proposition}
\begin{proof}
First we show that
for all
$J\subset S$,
\[
P(D_J^*)\colon \capker P_{n+1}\mapsto \capker P_{n+1}. 
\]
It can be easily obtained from the equality
\[
P_{n+1}P(D_J)^*=P(D_J)P(W_J)P(D_J)^*=P(D_J)P_{n+1}.
\]
Then by (\ref{adeul}), we have
     \[
U_n-(-1)^{n}1
\colon \capker P_{n+1}\mapsto \capker P_{n+1}.
\settowidth{\qedskip}{$\displaystyle
U_n-(-1)^{n}1
\colon \capker P_{n+1}\mapsto \capker P_{n+1}.$}
\addtolength{\qedskip}{-\textwidth}
\rlap{\hskip-0.5\qedskip\llap{\qedsymbol}}
     \]
\renewcommand{\qed}{\relax}
\end{proof}
\begin{proposition}
Let operators $\{T_i,\ i=1,\dots,n\}$ satisfy the braid condition
$T_iT_{i+1}T_i=T_{i+1}T_iT_{i+1}$, $i=1,\dots,n-1$, and
$T_iT_j=T_jT_i$, $\left| i-j\right|\ge 2$. Then
\begin{equation}\label{comrule}
     T_k U_n =U_n T_{n+1-k},\quad \forall\, k=1,\dots n.
\end{equation}
\end{proposition}
\begin{proof}
1.
For $n=1$ the equality is evident.

2.
Suppose that (\ref{comrule}) holds for any $n\le m$.
Note
that
\[
U_{m+1}=T_1T_2\cdots T_{m+1}U_m.
\]
Then,
for $1<k\le m+1$, we
have
\begin{align*}
T_k U_{m+1}& = T_k(T_1T_2\cdots T_{m+1})U_m \\
& =T_1T_2\cdots T_{k-2}T_kT_{k-1}T_kT_{k+1}\cdots T_{m+1}U_m\\
& =T_1T_2\cdots T_{k-2}T_{k-1}T_kT_{k-1}T_{k+1}\cdots T_{m+1}U_m\\
& =(T_1T_2\cdots T_{m+1}) T_{k-1}U_m\\
& =T_1T_2\cdots T_{m+1}U_m T_{m+1-(k-1)}\\
& =U_{m+1} T_{m+2-k}.
\end{align*}
In particular,
for $k=m+1$ we have $T_{m+1}U_{m+1}=U_{m+1}T_1$. Then taking
adjoints, we obtain the required equality for $k=1$.
\end{proof}

\noindent
{\bf 3.} Now we can formulate the main result of this paper.

\begin{theorem}
\label{thm:kersum}
Let $W(T)$ be a Wick algebra with braided operator
$T$ satisfying the norm bound $\Vert T\Vert\le 1$.
Then for any $n\ge 2$, we have
\[
\ker P_{n+1}=\sum_{k+l=n-1}
\hb^{\otimes\,k}\otimes \ker(1+T)\otimes\hb^{\otimes\,l}=
\sum_{k=1}^{n} \ker (1+T_k).
\] 
\end{theorem}
\begin{proof}
In fact, we shall prove the following:
Let $T_1, T_2,\dots ,T_n\in B(\mathcal{K})$, where $\mathcal{K}$ is
a
finite-dimensional Hilbert space, be selfadjoint contractions satisfying
the
relations
\[
T_iT_{i+1}T_i=T_{i+1}T_iT_{i+1},\ 1\le i\le n-1,\quad T_iT_j=T_jT_i
,\ \left| i-j\right|\ge 2.
\]
Then
\begin{equation}\label{eq:res}
\ker P_{n+1}=\smash[t]{\sum_{k=1}^n} \ker (1+T_k).
\end{equation}
(It
follows
trivially
from the decomposition
$P_{n+1}=P(D_{\{k\}})(1+T_k)$ that
$\sum_{k=1}^n \ker (1+T_k)\subset \ker P_{n+1}$).

We proceed using induction. 
\medskip

\noindent
\textbf{The case $n=2$.} 

In this case $P_2=1+T$.
\medskip

\noindent
\textbf{The case $n\mapsto n+1$.} 

It follows from $P_{n+1}=P(W_J)P(D_J)^*$ that 
\[
P(D_J)^*\colon
\ker P_{n+1}\mapsto
\ker P(W_J), 
\]
i.e., $\img  (P(D_J)^*|_{\ker P_{n+1}})\subset\ker P(W_J)$.
Moreover, it is obvious that for any $J\subset S$,
$\ker P(W_J)\subset \ker P_{n+1}$.
Therefore, by
(\ref{adeul}), we have the following inclusion:
\[
\img (U_n-(-1)^{n}1)|_{\ker P_{n+1}}\subset
\sum_{J\subset S,\,J\neq\emptyset}\ker P(W_J).
\]
Since, for $J\subset S$,
the group $W_J=W_{J_1}\times\cdots\times W_{J_k}$,
where $W_{J_l}\simeq S_{n_l}$ with $n_l<n+1$, we have a
decomposition into the product of pairwise commuting selfadjoint
operators
\[
P(W_J)=P(W_{J_1})\cdots P(W_{J_k}).
\]
Therefore
\[
\ker P(W_J)=\smash[t]{\sum_{l=1}^k}
\ker P(W_{J_l})\subset\smash[t]{\sum_{i=1}^{n}}
\ker(1+T_i),
\]
where the last inclusion is obtained from the assumption of induction.
So,
\begin{equation}\label{imbb}
\img (U_n-(-1)^{n}1)|_{\ker P_{n+1}}
\subset \sum_{i=1}^n\ker (1+T_i).
\end{equation}

Consider the operator $1-U_n^2$. Since $U_n=U_n^*\colon\ker
P_{n+1}\mapsto\ker P_{n+1}$, then
\[
\ker P_{n+1}=\img (1-U_n^2)|_{\ker P_{n+1}}+\ker(1-U_n^2)
|_{\ker P_{n+1}}.
\]
Moreover, since $\img (1-U_n^2)\subset \img (U_n-(-1)^n1)$,
using (\ref{imbb}), we have the inclusion
\[
\ker P_{n+1}\subset\smash[t]{\sum_{k=1}^n} \ker (1+T_i) +
\ker (1-U_n^2)|_{\ker P_{n+1}}.
\]
To finish the proof it remains only to show that
\[
\ker (1-U_n^2) \cap\ker P_{n+1}\subset\sum_{i=1}^n \ker (1+T_i).
\]
To this end,
we
may
present $1-U_n^2$ in the form
\begin{align*}
1-U_n^2  & = 1-T_1T_2\cdots T_n U_{n-1}^2 T_n\cdots T_2T_1\\
& = (1-T_1^2)+T_1(1-T_2^2)T_1 \\
& \quad +\cdots \\
& \quad + T_1T_2\cdots T_{n-1}(1-T_n^2)T_{n-1}\cdots T_2T_1 \\
& \quad + T_1T_2\cdots T_n (1-U_{n-1}^2)T_n\cdots T_2T_1.
\end{align*}
Since $\Vert T\Vert\le 1$ implies that $\Vert T_i\Vert\le 1 $,
$i=1,\dots n$, and $\Vert U_k\Vert\le 1$, $k\ge 2$, then we have
a
sum of
non-negative operators, and $v\in\ker (1-U_n^2)$ implies
that
\begin{align*}
& T_1^2 v=v,  \\
& T_2^2 T_1v=T_1v,\\
 &{}\mkern5mu\vdots \\
& T_n^2 T_{n-1}\cdots T_2T_1v=T_{n-1}\cdots T_2T_1v.
\end{align*}
However, $T_k U_n=U_nT_{n+1-k}$ implies that $T_k U_n^2=U_n^2T_k$,
and,
consequently,
\[
T_k\colon\ker (1-U_n^2)\mapsto\ker(1-U_n^2),\quad k=1,\dots,n.
\]
Moreover, since the restriction of $T_1$
to
$\ker(1- U_n^2)$ is
an
involution, 
\[
\img  (T_1)|_{\ker(1-U_n^2)}=\ker (1-U_n^2),
\]
and, for any
$v\in\ker(1-U_n^2)$, we have $T_2^2v=v$. By the same arguments, we
obtain that
\[
\forall\, v\in\ker (1-U_n^2),\quad T_i^2 v=v,\ i=1,\dots,d.
\]
Let now $v\in\ker (1- U_n^2)\cap\ker P_{n+1}$; then, for any
$k=1,\dots,n$,
\begin{align*}
P_{n+1}T_k v & =P(D_{\{k\}})(1+T_k)T_kv \\
 & =P(D_{\{k\}})(T_k+T_k^2)v  \\
 & =P(D_{\{k\}})(1+T_k)v =P_{n+1}v=0.
\end{align*}
Therefore $T_k$ maps $\ker(1-U_n^2)\cap \capker
P_{n+1}$ onto itself for any $k=1,\dots,n$. This fact
implies
that, for  any 
$\sigma\in S_{n+1}$,
we have $\phi(\sigma)v\in\ker(1-U_n^2)\cap\ker P_{n+1}$, 
and
\[
\forall\,k=1,\dots,n,\;\forall\,\sigma\in S_{n+1},\qquad
(1-T_k^2)\phi(\sigma)v=0.
\]
For
convenience, we
fix
the set $S_{n+1}$, and
set $v_i:=\phi(\pi_i)v$
for $\pi_i\in
S_{n+1}$ ($\pi_1:=\id$ and $v_1=v$).
Then
the
condition $P_{n+1}v=0$ takes the form
\begin{equation}\label{zer}
\sum_{k=1}^{n!}v_k=0.
\end{equation}
Finally, for any pair $i\ne j$ there exist generators
$\sigma_{i_1},\dots,\sigma_{i_m}\in S$ such that
\[
\pi_j=\sigma_{i_1}\cdots\sigma_{i_m}\pi_i
\]
and
\[
v_j = T_{i_1}\cdots T_{i_m}v_i.
\]
Note
that, if $v_k= T_r v_l$ for some $r=1,\dots,n$, then
$T_r^2v_k=v_k$ implies that $v_k-v_l\in\ker(1+T_r)$. Therefore,
for any
$i\ne j$,
\[
v_i-v_j\in\sum_{k=1}^n \ker(1+T_k).
\]
In particular,
for any $j=2,\dots,n!$, 
\[
v_1-v_j\in\smash[b]{\sum_{k=1}^n} \ker(1+T_k).
\]
Then from (\ref{zer}), we have
\[
n!\, v = n!\, v_1 \in\smash{\sum_{k=1}^n}\ker(1+T_k),
\]
and therefore
\[ 
v\in\smash[t]{\sum_{k=1}^n}\ker(1+T_k).
\settowidth{\qedskip}{$\displaystyle
v\in\smash[t]{\sum_{k=1}^n}\ker(1+T_k).$}
\addtolength{\qedskip}{-\textwidth}
\rlap{\hskip-0.5\qedskip\llap{\qedsymbol}}
\]
\renewcommand{\qed}{\relax}
\end{proof}
\begin{remark}
Evidently the proof does not depend on the dimension of $\mathcal{K}$.
Indeed, in the case
when
$\mathcal{K}$ is infinite-dimensional, the
linear subspace in (\ref{eq:res}) is replaced by its closure. I.e., if
$\mathcal{K}$ is
a
separable Hilbert space and
$\{T_i,\ i=1,\dots, n\}$ are selfadjoint contractions satisfying the
braid conditions, then
\[
\ker P_{n+1}=\overline{\sum_{k=1}^n\ker (1+T_k)}.
\]
\end{remark}
As a corollary we have
an
improved version of the result of
Bo\.zejko and Speicher (see \cite{bib:2}).
\begin{proposition} \label{ff}
If
the
operator $T$ satisfies the braid condition, and $-1< T\le 1$, then 
$P_n>0$, $n\ge 2$, i.e., the Fock inner product is strictly positive,
and the
Fock representation acts in the whole space $\mathcal{T}(\hb)$.
\end{proposition}
\begin{proof}
Recall that if $T$ is braided and $\Vert T\Vert\le 1$ then $P_n\ge 0$
(see \cite{bib:2}). It remains only to show that
$\ker P_n=\{0\}$
for $-1<T\le 1$.
This fact trivially follows from our theorem since in this case
$\ker (1+T)=\{0\}$.
\end{proof}
\section{Corollaries and examples}\label{S3}
We summarize
the
results obtained above in the following proposition.
\begin{proposition}\label{wi}
If $W(T)$ is a Wick algebra with braided operator of coefficients $T$
satisfying the norm bound $\Vert T\Vert\le 1$, then
the following three statements hold.
\begin{enumerate}
\item The kernel of
the
Fock
representation is generated by the largest quadratic Wick ideal.
In particular,
if $-1<T\le 1$, then the Fock representation is faithful.
\item For any $n\ge 2$ we have
the
inclusion
$\mathcal{I}_n\subset\mathcal{I}_2$.
\item If $-1<T\le 1$, then $W(T)$ has no
non-trivial Wick ideals.
\end{enumerate}
\end{proposition}
\begin{example}
Consider the $q$-CCR algebra
based on a Hilbert space $\hb$ and the relations
\begin{align*}
a_i^*a_i & =1+qa_ia_i^*,\quad i=1,\dots,d,\\
a_i^*a_j & =qa_ja_i^*,\quad i\ne j,\quad 0<q<1.
\end{align*}
We pick an orthogonal basis $(e_i)$ in $\hb$, and then
$T$ is determined on
this
basis by
the
formulas
\[
T e_i\otimes e_j = q e_j\otimes e_i,\quad \Vert T\Vert<1.
\]
It is evident that $T$ is braided. Then by
the
proposition,
we cannot have any
Wick ideals in $W(T)$.
\end{example}

It was proved in \cite{bib:2} that for braided $T$ satisfying the norm bound
$\Vert T\Vert<1$,
the Fock representation is bounded. Therefore we may
consider the $C^*$-algebra generated by operators of
the
Fock
representation.

Recall that a $*$-algebra is called $C^*$-representable if it can
be realized as a $*$-subalgebra of a certain $C^*$-algebra (see for
example \cite{bib:9}).
Combining
the
results of Theorem~\ref{thm:kersum}
and Proposition~\ref{ff}, we obtain the
following statement.
\begin{proposition}
If $W(T)$ is a Wick algebra with braided operator of coefficients $T$
satisfying the norm bound $\Vert T\Vert < 1$, then $W(T)$ is
$C^*$-representable.
\end{proposition}
Suppose that, in the case of braided $T$ with $\Vert T\Vert= 1$ and
$\ker (1+T)\ne \{0\}$, the Fock representation is bounded. Then
Theorem~\ref{thm:kersum} implies that
the
quotient $W(T)/\mathcal{I}_2$ is $C^*$-representable.
\begin{example}
Consider the following type of $q_{ij}$-CCR (see \cite{bib:2}):
\begin{align*}
a_i^*a_i& =1+q_i a_ia_i^*,\quad i=1,\dots,d,\quad 0<q_i<1,\\
a_i^*a_j& =\lambda_{ij}a_ja_i^*,\quad i\ne j,\ \left|\lambda_{ij}\right|=1,
\ \lambda_{ij}=\overline{\lambda}_{ij}.
\end{align*}
The
corresponding
$T$ is braided, $\Vert T\Vert=1$, and
\[
\ker (1+T) =\bigl< a_ja_i-\lambda_{ij}a_ia_j,\ i<j\bigr>.
\]
Moreover, the Fock representation of this algebra is bounded.
Then, as noted above, the $*$-algebra generated by
the
relations
\begin{align*}
a_i^*a_i& =1+q_i a_ia_i^*,\quad i=1,\dots,d,\quad 0<q_i<1,\\
a_i^*a_j& =\lambda_{ij}a_ja_i^*,\quad i\ne j,\ \left|\lambda_{ij}\right|=1,
\ \lambda_{ij}=\overline{\lambda}_{ij},\\
a_j a_i &=\lambda_{ij}a_ia_j,\quad i<j
\end{align*}
is $C^*$-representable.

A
description of
the
irreducible representations of
these
relations can
be found for example in \cite[sec.~2.4]{bib:11}.
\end{example}

Note that, if $\Vert T\Vert = 1$, then the operators of
the
Fock representation
can be unbounded.
\begin{example}
Consider the following Wick algebra:
\begin{align*}
a_i^*a_i& = 1 + a_i a_i^*,\quad i=1,\dots,d,\\
a_i^*a_j& = q a_j a_i^*,\quad i\ne j,\quad -1<q<1.
\end{align*}
The
corresponding
$T$ is determined by the formulas
\[
T e_i\otimes e_i = e_i\otimes e_i,\quad
T e_j\otimes e_i = q e_i\otimes e_j,\ i\neq j,\ i=1,\dots,d.
\]
It is easy to see that $T$ is braided and $-1 < T\le 1$. So, the Fock
representation of this algebra is faithful. Note that, if we consider
the complement of $\mathcal{T}(\hb)$ with respect to the Fock inner
product, then
the
operators of
the
Fock representation are unbounded.

For
the definition and properties of
representations of
$*$-algebras
by unbounded operators, see for example
\cite{bib:10}.

Unbounded representations of Wick algebras will be
considered in more detail later.
\end{example}

\noindent
{\bf Acknowledgements.}\ \ 
Yu.\ Samo\u{\i}lenko and Dan.\ Proskurin express their gratitude to
Prof.\ Vasyl L. Ostrovsky\u{\i} and Stanislav Popovych for their attention
and helpful discussions.
We also thank Brian Treadway for proofreading and polishing.


\begin{thebibliography}{99}

\bibitem{bib:4} { N. Bourbaki}, \textit{%
El\'ements de mathematique, Fasc.\ XXXIV: Groupes et algebres de Lie},
Chap.\ 4, 5, 6,
Hermann, Paris, 1968.

\bibitem{bib:2}
{ M. Bo\.zejko and R. Speicher}, \textit{%
Completely positive maps on
Coxeter groups, deformed commutation relations, and operator
spaces}, Math.\ Ann.\ \textbf{300} (1994), 97--120.

\bibitem{bib:7} { K. Dykema and A. Nica}, \textit{%
On the Fock representation
of the $q$-commutation relations}, J.~Reine Angew.\ Math.\ 
\textbf{440} (1993), 201--212.

\bibitem{bib:5} { O.W. Greenberg}, \textit{%
Particles with small violations of
Fermi or Bose statistics}, Phys.\ Rev.~D (3) \textbf{43} (1991),
4111--4120.

\bibitem{bib:8} { P.E.T J{\o}rgensen, L.M. Schmitt and R.F. Werner},
\textit{%
$q$-canonical commutation relations and stability of the Cuntz algebra},
Pacific J. Math.\ \textbf{163} (1994), no.~1, 131--151.

\bibitem{bib:1} { P.E.T. J{\o}rgensen, L.M. Schmitt, and R.F. Werner},
\textit{%
Positive representations of general commutation relations allowing Wick
ordering},  J.~Funct.\ Anal.\ {\bf 134} (1995), 33--99.

\bibitem{bib:9} { E.C. Lance}, \textit{%
Finitely-presented $C\sp *$-algebras},
Operator Algebras and Applications,
NATO Adv.\ Sci.\ Inst.\ Ser.\ C Math.\ Phys.\ Sci., vol.~495,
Kluwer Acad.\ Publ.,
Dordrecht,  1997,
pp.~255--266.

\bibitem{bib:11} {V. Ostrovsky\u\i{} and Yu.\ Samo\u\i{}lenko}, \textit{%
Introduction
to the Theory of Representations of Finitely Presented $*$-Algebras, I:
Representations by Bounded Operators}, The Gordon and Breach
Publishing Group, London, 1999.

\bibitem{bib:6} { D. Proskurin}, \textit{%
Homogeneous ideals in Wick $*$-algebras},
Proc.\ Amer.\ Math.\ Soc.\ \textbf{126} (1998), no.~11, 3371--3376.

\bibitem{bib:3} { W. Pusz and S.L. Woronowicz}, 
\textit{Twisted second quantization},
Rep.\ Math.\ Phys.\ {\bf 27} (1989), 231--257.

\bibitem{bib:10} { K. Schm\"u{}dgen}, \textit{%
Unbounded Operator Algebras and
Representation Theory},
Operator Theory: Advances and Applications, vol.~37,
Birkh\"auser, Basel, 1990.

\end{thebibliography}
\end{document}